\documentclass[doublecol]{epl2}

\title{The double life of electrons in magnetic iron pnictides, as revealed by NMR}
\shorttitle{Double life of electrons in magnetic iron pnictides} 

\author{Andrew Smerald\inst{1} \and Nic Shannon\inst{1}}
\shortauthor{A. Smerald and N. Shannon}

\institute{                    
  \inst{1} H.\ H.\ Wills Physics Laboratory, University of Bristol,  Tyndall Av, BS8--1TL, UK.
}

\abstract{
We present a phenomenological, two-fluid approach to understanding the magnetic excitations in Fe 
pnictides, in which a paramagnetic fluid with gapless, incoherent particle-hole excitations coexists with 
an antiferromagnetic fluid with gapped, coherent spin wave excitations.   We show that this two-fluid 
phenomenology provides an excellent quantitative description of NMR data for magnetic ``122'' pnictides, 
and argue that it finds a natural justification in LSDA and spin density wave calculations.    We further 
use this phenomenology to estimate the maximum renormalisation of the ordered moment that can 
follow from low-energy spin fluctuations in Fe pnictides.   We find that this is too small to account for 
the discrepancy between {\it ab intio} calculations and 
neutron scattering measurements.}

\begin{document}

\maketitle


\section{Introduction}

The discovery that, suitably doped, Fe pnictides can superconduct at temperatures greater than 50K~\cite{kamihara08}
has sparked a sudden rush of interest in these materials.  As with the high-$T_c$ cuprates, the 
undoped parent compounds are magnetic.  Neutron scattering and $\mu$SR experiments suggest a direct 
competition between the two states, with the magnetism winning at low doping and the superconductivity 
taking over as the doping is increased\cite{lester09,uemura10,wilson10}.    Understanding the magnetic excitations 
in these materials is therefore widely believed to be an important step towards understanding their superconductivity, 
as well as an interesting problem in its own right.

To date, most theoretical approaches to this problem 
have stressed either the itinerant nature of electrons in Fe pnictides\cite{han09,yaresko09,yi09}, or used strong electronic 
correlation to justify mapping them onto a frustrated local moment model\cite{si08,yao08,uhrig09}.   
In this paper we embrace the fact that Fe pnictides are both metals and magnets, proposing a simple, 
phenomenological, two-fluid description of their magnetic excitations.  
We argue that spin excitations at low energies and temperatures are dominated by the gapless, incoherent 
particle-hole excitations, characteristic of a metallic paramagnet, while for energies and temperatures comparable 
with a spin gap $\Delta_\sigma$, coherent, collective excitations of the magnetic order come into play.    
It follows naturally from experimental and theoretical determinations of the band structure that these two fluids are essentially independent.  


While this two-fluid phenomenology is not tied to any particular microscopic model, it finds a natural justification in recent spin density wave (SDW) calculations\cite{korshunov08,klauss08,ying-ran09,kaneshita09,eremin10,knolle10,kaneshita-arXiv}, ARPES experiments \cite{yi09,hsieh09} and LSDA calculations \cite{yi09}. In this paper we further show that our phenomenology provides an excellent description of NMR experiments on Fe pnictides 
with 122 structure \cite{kitagawa08,kitagawa09}.

We go on to address a second major issue in the Pnictide materials, namely the role of frustration. It has been suggested that the large discrepancy in the size of the ordered moment between {\it ab intio} calculations and 
neutron scattering measurements can be understood by fine tuning a frustrated local moment model \cite{yao08,uhrig09,si08}. We critically re-examine such a model in terms of our
two-fluid phenomenolgy and conclude that, while it does not rule out frustration {\it per se}, 
quantum fluctuations {\it cannot} account for the observed reduction of the ordered moment 
relative to LDA calculations \cite{han09,yaresko09,yi09}.
 
  
Both the magnetic and metallic properties of Fe pnictides originate in outer-shell Fe 3d-electrons.   
Band structure calculations\cite{singh08,han09,yaresko09}, supported by photoemission\cite{liu08,evtushinsky09} 
and quantum oscillation\cite{coldea08,analytis09} experiments, suggest that these hybridize with As 4p orbitals 
to form a Fermi surface with two electron-like and three hole-like pockets, 
when viewed in a ``natural'' unfolded Brillouin zone based on Fe sites. 
SDW calculations \cite{korshunov08,klauss08,ying-ran09,kaneshita09,eremin10,knolle10,kaneshita-arXiv}, 
ARPES experiments \cite{yi09,hsieh09} and LSDA calculations \cite{yi09} find general agreement on a number of points. There are observed to be five bands crossing the Fermi surface in the paramagnetic state. These  undergo a non-trivial reconstruction at the magnetic ordering transition, with some of the bands mixing to form a gapped SDW state and the rest remaining metallic. Furthermore, there is no pair of metallic bands in the ordered state that is nested with spanning vector $(\pi,0)$.  
  For example, the calculations presented in \cite{eremin10} consider four of the five bands, two hole-like and two electron-like, and show that $(\pi,0)$ order arises most naturally when only one of the hole bands takes part in the Fermi surface mixing.  Therefore the magnetically ordered state retains a hole-like sheet of Fermi surface, which will support metallic, particle-hole excitations. 
Similarly,  by comparing ARPES experiments and LSDA calculations, Ref.~\cite{yi09} shows that below $T_{SDW}$ the Fermi surface reconstructs to form $(\pi,0)$ SDW order, but ungapped, metallic Fermi surface pockets remain, centred on the $\Gamma$ point.

The fact that magnetic Fe pnictides are metals implies that some part of this complex 
Fermi surface remains gapless, and will support incoherent particle-hole excitations with 
vanishing energy.     We treat this as the first of our fluids, characterised simply by an average density 
of states at the Fermi energy, $n_0$, which can be estimated from heat capacity measurements.
 
  
Neutron scattering experiments\cite{zhao08,zhao09,diallo09,matan09}, meanwhile, reveal a commensurate, collinear, antiferromagnetic (AF)
ground state with ordering vector ${\bf k^*}=(\pi,0,\pi)$, and ordered Fe moment $m_S \approx 1 \mu_B$, 
much smaller than predicted by {\it ab initio} calculations\cite{han09,yaresko09}.  
A single branch of spin wave excitations with dispersion, 
\begin{eqnarray}
\label{eq:cone}
\omega_{\bf k^\prime}=\sqrt{\Delta_\sigma^2+v_x^2k_x^2+v_y^2k_y^2+v_z^2k_z^2}, 
\end{eqnarray}
is found above a gap $\Delta_\sigma \approx 10 meV$ at the ordering vector ${\bf k^\prime} = {\bf k} - {\bf k^*} = (0,0,0)$.
Spin wave velocities \mbox{${\bf v}=(v_x,v_y,v_z)$} are anisotropic, with $v_x \approx v_y \gg v_z$.  
The collective excitations of this magnetic order form our second fluid, and, following \cite{ong09},
 we characterise 
them using a quantum non-linear sigma model, 
 \begin{eqnarray}
\label{eq:S} 
S &=&\frac{1}{2abc} \int d{\bf x} dt 
\left[  
\hbar^2\chi_\perp (\partial_t {\bf n})^2 - \rho_x (\partial_x {\bf n})^2 - \rho_y (\partial_y {\bf n})^2 
\right. 
\nonumber\\
&& \qquad\qquad\qquad
 \left. - \rho_z (\partial_z {\bf n})^2
+\chi_\perp \Delta_\sigma^2 n_x^2  
 \right],
\end{eqnarray}
where $\chi_\perp$ is the static perpendicular susceptibility, $\rho_x, \rho_y$ and $\rho_z$ are spin stiffness' along the 
Fe-Fe crystal axes $a,b,c$, and $\Delta_\sigma^2$ is an easy axis anisotropy.   


For $\Delta_\sigma \to 0$, this action describes the long-wavelength Goldstone modes, which follow from the symmetry of the magnetic order.   
As such, it can be derived from {\it any} microscopic model that respects these symmetries, whether localised or itinerant.   
For finite anisotropy $\Delta_\sigma > 0$, Eq.~(\ref{eq:S}) predicts a gapped, two-fold degenerate cone of spin wave excitations with 
exactly the form of Eq.~(\ref{eq:cone}), where \mbox{$v_\alpha = \sqrt{\rho_\alpha/\chi_\perp}$}.  

For collinear order it is natural to consider a $\mathcal{Z}_2$ symmetry~\cite{chandra90,lante06} between $(\pi,0)$ and $(0,\pi)$ states, which is not encoded in the non-linear sigma model. However, in the Pnictides, this symmetry is broken by a tetragonal to orthorhombic phase transition that typically occurs at or very close to the magnetic ordering temperature~\cite{jesche08}. Below this temperature the simpler non-linear sigma model description is sufficient, provided there are well defined cones of spin wave excitations.

Within a spin density wave picture, Eq.~(\ref{eq:S}) should remain valid up to an energy scale of the spin-density wave 
gap, estimated to be $\Delta_{\sf SDW} \approx 31 meV$ for LaFeAsO\cite{knolle10}.
For the specific case of $\mathrm{CaFe_2As_2}$, it breaks down at energies of approximately 150meV,
where the spin wave branch is seen to enter a continuum of excitations \cite{zhao09}.  


Our final approximation is to ignore all coupling between these two fluids.   
This appears justified for two reasons.   
Firstly neutron scattering studies\cite{matan09,diallo09,zhao09} observe sharp cones of low energy magnetic excitations with no evidence of the damping that would be expected if the spin waves could scatter from the metallic fluid.     
Secondly LSDA calculations\cite{yi09}, photoemission studies\cite{yi09} and SDW theory\cite{korshunov08,klauss08,ying-ran09,kaneshita09,eremin10,knolle10,kaneshita-arXiv} show no evidence, 
in the magnetic ordered state, for a nested pair of Fermi surfaces with spanning vector $(\pi,0)$.  Hence, within a band picture, there are no available particle-hole states close to the ordering vector for the spin waves to decay into.

We note that there is evidence \cite{ying-ran09,yi09} for a node in the SDW gap. However, this is not situated at the ordering vector, is not nested with any other Fermi surface with spanning vector $(\pi,0)$ and has vanishing density of states. Thus it does not appear relevant to our low temperature model.

\section{Sublattice magnetisation}

\begin{figure}[ht]
\centering
\onefigure[width=0.48\textwidth]{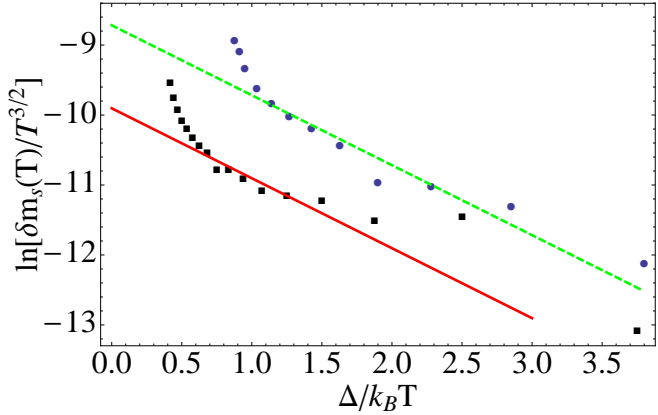}
\caption{\footnotesize{(Colour online).  Temperature dependence of the ordered moment $\delta m_S(T)$ as determined by NMR 
measurements on $\mathrm{BaFe_2As}_2$~\cite{kitagawa08} (blue circles) and $\mathrm{SrFe_2As}_2$~\cite{kitagawa09} (black squares).    
Data is plotted as $\ln\left[ \delta m_S(T)/T^{\frac{3}{2}}\right]$ vs $\Delta_\sigma/k_BT$, where the values of 
$\Delta_{Ba}=114K$ and $\Delta_{Sr}=75K$ are taken from inelastic neutron scattering experiments\cite{matan09,zhao08}.   
Straight lines show the expected form of corrections at low temperatures. The intercept gives the prefactors $M_{\sf LT}^{Ba} \approx 1.6 \times 10^{-4}  \ \mu_B K^{-\frac{3}{2}}$ and $  M_{\sf LT}^{Sr} \approx 5 \times 10^{-5} \ \mu_B K^{-\frac{3}{2}}$}}
\label{fig:logmag}
\end{figure}

A simple test of our phenomenology is  provided by the temperature dependence of the ordered moment $\delta m_S(T)$.  This renormalises the zero temperature moment according to $m_s=m_0-\delta m_S-\delta m_S(T) $, where $m_0$ is the bare moment and $\delta m_S$ describes the reduction due to quantum fluctuations.
Within the two-fluid picture, $\delta m_S(T)$ is controlled by the thermal excitation of spin waves, as described by Eq.~(\ref{eq:S}).  
For $T \ll \Delta_\sigma$ we find activated behaviour,
 \begin{eqnarray}
\label{eq:mslowT}
\delta m_S(T)
&=& \frac{m_0(abc)\sqrt{\Delta_\sigma}}{8\chi_\perp\bar{v}_s^3} \left( \frac{2k_BT}{\pi}\right)^{\frac{3}{2}} \mathrm{e}^{-\frac{\Delta_\sigma}{k_BT}}  \nonumber \\
&=& M_{\sf LT} T^{\frac{3}{2}}\mathrm{e}^{-\frac{\Delta_\sigma}{k_BT}},
\end{eqnarray}
while for $T \gg \Delta_\sigma$ we find the power law behaviour,
 \begin{eqnarray}
\label{eq:mshighT}
\delta m_S(T)
=\frac{m_0(abc)}{12\chi_\perp\bar{v}_s^3}(k_BT)^2  
= M_{\sf HT}T^2.
\end{eqnarray}
The form of corrections depends only on the gap, $\Delta_\sigma$.   
The prefactor is determined by the geometric mean spin wave velocities,
$\bar{v}_{s}^3=v_xv_yv_z \ [(meV\AA)^3]$, bare moment $m_0 \ [\mu_B]$,  transverse susceptibility $\chi_\perp \ [meV^{-1}]$ and Fe-Fe lattice parameters $a,b,c \ [\AA]$.
At temperatures relevant to experiment, the spin gap dominates, and, 
in Fig.~\ref{fig:logmag}, we compare the predicted form of $\delta m_S(T)$ with the low temperature ordered moment,  
as measured by NMR experiments on $\mathrm{BaFe_2As}_2$~\cite{kitagawa08} and 
$\mathrm{SrFe_2As}_2$~\cite{kitagawa09}.     
We have checked that similar fits can be made for $\delta m_S(T)$ obtained from 
$\mu$SR for $\mathrm{LaOFeAs}$~\cite{carlo09,uemura09} and $\mathrm{SrFe_2As}_2$~\cite{jesche08}.


\section{Spin-lattice relaxation rate}

NMR experiments also probe spin excitations through the nuclear spin lattice relaxation rate, $1/T_1$.
This has been measured for As nuclei in $\mathrm{BaFe_2As}_2$~\cite{kitagawa08} and $\mathrm{SrFe_2As}_2$~\cite{kitagawa09}.
For hyperfine interactions,  
the relaxation rate is given by\cite{moriya56,moriya63},
\begin{eqnarray}
\frac{1}{T_1} \approx \frac{\gamma_N^2}{2} k_BT \lim_{\omega_0 \to 0} \frac{1}{N} \sum_{{\bf q}}  
|\mathcal{A}_{\bf q}|^2 
\frac{ \chi^{\prime\prime}(\omega_0,{\bf q}) }{\hbar \omega_0},
\end{eqnarray}
where $\gamma_N$ is the gyromagnetic ratio of the nucleus in question, $\omega_0$ is the nuclear excitation energy, $|\mathcal{A}_{\bf q}|^2$ is a form factor describing the  coupling between the surrounding electrons and the nuclear spin and $ \chi^{\prime\prime}(\omega_0,{\bf q})$ is the imaginary part of the longitudinal, dynamic susceptibility of the electron system.
Both fluids contribute to $1/T_1$, but at low temperatures the leading contribution will come
from gapless particle-hole pairs within the paramagnetic fluid.       
We assume a roughly constant contact interaction between the nucleus and the metallic electrons, $|\mathcal{A}_{\bf q}|^2 \approx |\mathcal{A}_0|^2  \ [(T/\mu_B)^2]$, 
over the relevant sheets of the Fermi surface. This leads to a contribution to $1/T_1$ which is linear 
in $T$~\cite{moriya63},
\begin{eqnarray}
\label{eq:T1inc}
1/T^{\sf inc.}_1 \approx \frac{1}{4} \pi \hbar \gamma_N^2 |\mathcal{A}_0 |^2 g_L^2(abc)^2 n_0^2  k_B T = C_{inc}T,
\end{eqnarray}
where $g_L$ is the Land\'{e} g-factor, $n_0$ is the density of states at the Fermi surface and  we use units where $\mu_B=1$.

\begin{figure*}[ht]
\centering
\onefigure[width=0.7\textwidth]{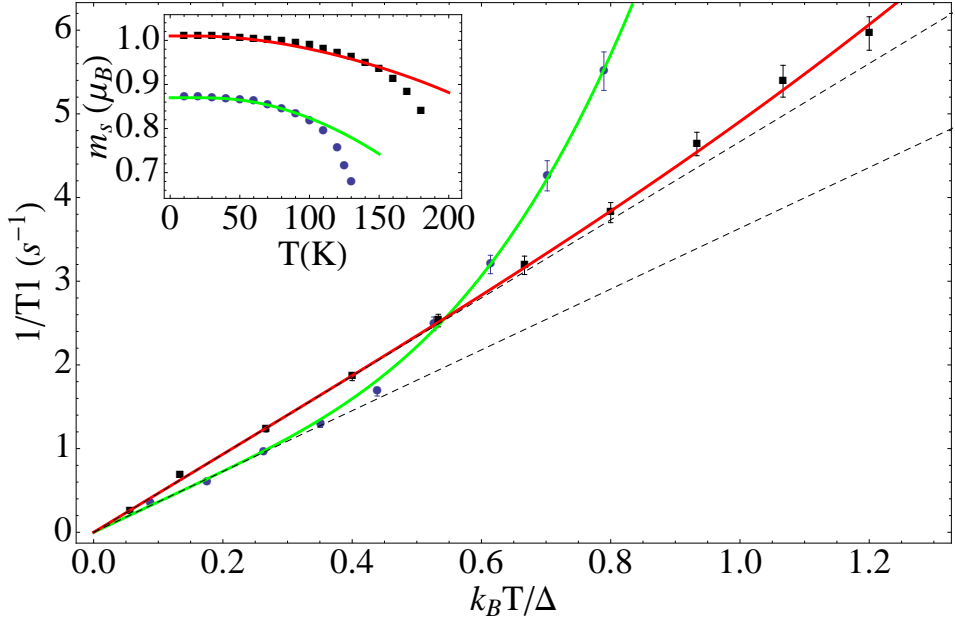}
\caption{\footnotesize{(Colour online).  Simultaneous fits to nuclear relaxation rate $T_1^{-1}$  and sublattice magnetsation 
$m_S(T)$ for $\mathrm{BaFe_2As}_2$\cite{kitagawa08} (blue dots) and $\mathrm{SrFe_2As}_2$\cite{kitagawa09} (black squares). The external field is applied in the $(1,1,0)$ direction.
The dashed lines show the contribution of incoherent particle-hole pairs
Eq.~(\ref{eq:T1inc}); 
the full lines show the combined fit including the contribution of coherent, thermally-activated spin waves Eq.~(\ref{eq:T1coh}). 
The gap values 
$\Delta_{Ba}=114K$ and $\Delta_{Sr}=75K$ are taken from inelastic neutron scattering experiments\cite{matan09,zhao08}. 
Insets show simultaneous fits to the sublattice magnetisation, $m_S$. The prefactors are determined to be, $C^{Ba}_{inc} \approx 0.032 \ s^{-1}K^{-1}$, $C^{Ba}_{coh} \approx 7.5 \ s^{-1}$, $C^{Sr}_{inc} \approx 0.062 \ s^{-1} K^{-1}$ and $C^{Sr}_{coh} \approx 0.28 \ s^{-1}$.
}}
\label{fig:T1}
\end{figure*}


\begin{figure}[ht]
\centering
\onefigure[width=0.35\textwidth]{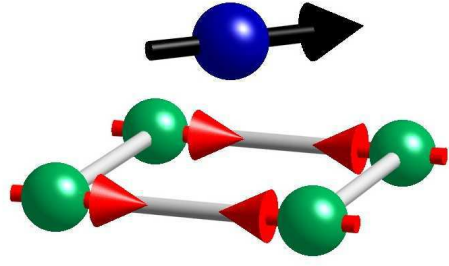}
\caption{\footnotesize{(Colour online). The local environment of the nuclear As spin (central blue atom) considered in the calculation of $1/T_1$ is a four site plaquette of iron spins (red arrows). The As atom lies in the centre of the rectangular plaquette but shifted out of the plane.
}}
\label{fig:Asenvironment}
\end{figure}

At higher temperatures, the Raman scattering of thermally excited spin waves also plays a role in nuclear
spin relaxation. This is dominant over single spin wave excitations since $\Delta_\sigma \gg \hbar\omega_0$.   
NMR probes the longitudinal susceptibility $\chi^{\prime\prime}_\parallel(\omega_0,{\bf q})$, which can be calculated directly from Eq.~(\ref{eq:S}).  
We consider relaxation due to coupling of the As nucleus to a four site plaquette of nearest neighbour Fe sites, 
shown in Fig.~\ref{fig:Asenvironment}, in the same spirit as for the Y nucleus in YBa$_2$Cu$_3$O$_{6+x}$ in \cite{mila89}.

In both cases, the antiferromagnetically ordered electron moments create an internal field at the nuclear site, and it is this that dictates 
the behaviour of the form factor, $ |\mathcal{A}_{\bf q}|^2$.
In the case of YBa$_2$Cu$_3$O$_{6+x}$ the combined symmetry of the crystal structure and magnetic order 
causes the internal field at the Y-site to disappear, and spin fluctuations are filtered by a form factor which 
vanishes at the magnetic ordering vector \cite{mila89}.
In contrast, the As site in a pnictide such as BaAs$_2$Fe$_2$ experiences a finite internal field directed along the c-axis~\cite{kitagawa08}, 
and for NMR fields applied in the $ab$-plane, longitudinal fluctuations of the ordered Fe moment couple efficiently to the nuclear spin.   
In this case, the appropriate form factor is
\begin{eqnarray}
 |\mathcal{A}_{\bf q}|^2=
 4\mathcal{B}_{ac}^2\left( 1-\cos q_x +\cos q_y -\cos q_x \cos q_y \right),
 \label{eq:formfactor}
\end{eqnarray}
where \mbox{$\mathcal{B}_{ac}$} is the matrix element relevant for Raman relaxation processes. 
The form factor and the imaginary part of the longitudinal susceptibility are simultaneously peaked at the ordering vector, ${\bf q}=(\pi,0,\pi)$.
The form factor is very slowly varying in comparison with the susceptibility, and hence we approximate it with the constant \mbox{$|\mathcal{A}_{\bf q}|^2 \approx 16\mathcal{B}_{ac}^2$}.  
The more complex case of external magnetic field parallel to the c-axis will be discussed elsewhere~\cite{inpreparation}. 

Making these approximations, we find,
\begin{eqnarray}
\label{eq:T1coh}
 \frac{1}{T^{\sf coh.}_1}  &\approx& 
 \frac{2\mathcal{B}_{ac}^2 m_0^2 \hbar (abc)^2 \gamma_N^2 \Delta_\sigma^3}{\pi^3 \chi_\perp^2\bar{v}_s^6}  
 \Phi \left[ \frac{k_B T}{\Delta_\sigma} \right] \nonumber \\
& \approx& C_{coh}  \Phi \left[ \frac{k_B T}{\Delta_\sigma} \right]
\end{eqnarray} 
where,
\begin{eqnarray} 
\Phi (x) = x^2 \mathrm{Li}_1(e^{-1/x}) + x^3 \mathrm{Li}_2(e^{-1/x}),
\end{eqnarray} 
and 
\mbox{$\mathrm{Li}_n(z) = \sum_{k=0}^\infty z^k/k^n$} is the n$^{th}$ polylogarithm of $z$.


%
\begin{table}
\begin{center}
\footnotesize
  \begin{tabular}{| c | c | c | c | c | }
    \hline
    & $M_{\sf LT}^{Ba} [\mu_B K^{-\frac{3}{2}}]$  & $C_{inc}^{Ba} [(sK)^{-1}]$ &  $C_{coh}^{Ba} [s^{-1}]$  \\ \hline 
Fit &  $1.6 \times 10^{-4}$ & 0.032  & 7.5  \\ \hline
Estimate & $(0.15-2.2) \times 10^{-4}$ & $\sim 0.015$ & $0.14-33$  \\ \hline
    \end{tabular}
\end{center} 
\caption{\footnotesize{Quantitative analysis of $\delta m_s(T)$ and $T_1^{-1}$  in BaFe$_2$As$_2$. 
We determine the prefactors $M_{\sf LT}^{Ba}$ (Eq.~\ref{eq:mslowT}), $C_{inc}^{Ba}$ (Eq.~\ref{eq:T1inc}),  
and $C_{coh}^{Ba}$ (Eq.~\ref{eq:T1coh}) by fitting NMR experiments (cf. Figs.~\ref{fig:logmag}, \ref{fig:T1}) and 
compare these with the quantitative estimates which follow from known values of the hyperfine interactions 
$\mathcal{A}_0\sim 1.88 \ T/\mu_B$ and $\mathcal{B}_{ac}=0.43 \ T/\mu_B$\cite{kitagawa08}, spin-wave velocities $95<\bar{v}_s<228 \ meV\AA \ $\cite{matan09}, 
spin-gap $\Delta_\sigma=9.8(4) \ meV \ $ \cite{matan09}, 
ordered moment $m_0=0.87 \ \mu_B$\cite{huang08}, perpendicular susceptibility $\chi_\perp=10^{-4} \ emu/mol \ $\cite{ning09}, lattice constants $[a,b,c]=[2.80,2.79,6.47]\ \AA \ $\cite{huang08}, 
and density of states $n_0=5.8\times 10^{24} \ J^{-1}m^{-3} \ $\cite{rotter08}.   
}}
\label{tab:prefactors}
\end{table}%
 
\section{Comparison to experiment}

We are now in a position to compare directly with experiment, and, in Fig.~\ref{fig:T1}, we show the results of simultaneous fits to NMR data 
for $\delta m_S(T)$ and $1/T_1$ with field in the $(1,1,0)$ direction in $\mathrm{BaFe_2As}_2$\cite{kitagawa08} and $\mathrm{SrFe_2As}_2$\cite{kitagawa09}.
We treat the total relaxation rate as the sum of  the contributions of the two fluids, Eq.~(\ref{eq:T1inc}) and Eq.~(\ref{eq:T1coh}) and 
fit the prefactors $M_{\sf LT}^{Ba}$ (Eq.~\ref{eq:mslowT}), $C_{inc}^{Ba}$ (Eq.~\ref{eq:T1inc}),  
and $C_{coh}^{Ba}$ (Eq.~\ref{eq:T1coh}), using the experimental value \mbox{$\Delta_{Ba}=114K$}\cite{matan09} 
for the gap.   The agreement for these two parameter fits is excellent. 

It is possible to make independent, {\it quantitative} estimates of these fit parameters by substituting known values
of the spin wave velocities, hyperfine interactions, density of states at the Fermi surface, Fe-Fe lattice parameters, perpendicular susceptibility and zero temperature sublattice magnetisation
directly into    
Eq.~(\ref{eq:T1inc}), Eq.~(\ref{eq:T1coh}) and  Eq.~(\ref{eq:mslowT}). 
In Table~\ref{tab:prefactors} we show that, within experimental error, this quantitative approach to the prefactors is consistent with the fits to NMR data
 for $\mathrm{BaFe_2As}_2$.
Uncertainty in the model parameters for $\mathrm{SrFe_2As}_2$ are currently too great for a quantitative comparison.

\section{The ordered moment at zero temperature}

\begin{figure*}[ht]
\centering
\onefigure[width=0.7\textwidth]{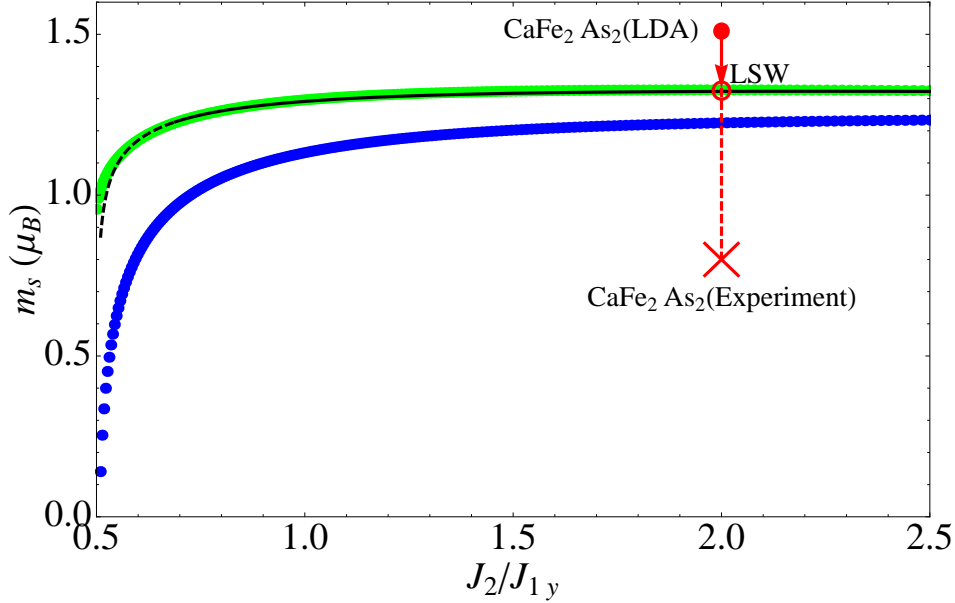}
\caption{\footnotesize{(Colour online).  Zero temperature sublattice magnetisation 
$m_S$ calculated within linear spin wave theory (LSW) for the 3D Heisenberg model Eq.~(\ref{eq:H})
(upper, green dots) and the square-lattice $J_1$--$J_2$ model (lower, blue dots), 
as a function of $J_2$.  
Remaining parameters for Eq.~\ref{eq:H} are taken from experiment on $\mathrm{CaFe_2As_2}$~\cite{diallo09}.   We use $m_0=1.51$, $J_{1x}=30meV$, $J_{1y}=15meV$, $J_{1z}=4.5meV$ and $K_{xy}=K_{z}=0.12meV$.
The solid black line shows the sigma-model prediction Eq.~(\ref{eq:mzp}).  
The divergent correction seen in the 2D $J_1$--$J_2$ model for \mbox{$J_2/J_1 \to^+ 1/2$} is 
cut off by the gap spin $\Delta_\sigma$ and 3D spin-wave dispersion. 
As a result the renormalisation of the bare moment (filled red circle) is insufficient to agree with the 
experimental value (red cross)\cite{diallo09}.
}}
\label{fig:mzp}
\end{figure*}


One of the important issues in Fe pnictide magnetism has been the size of the ordered moment $m_S$.  Fe and its oxides typically 
show a large ordered moment at low temperatures. First principles calculations for magnetic Fe pnictides suggest that 
$m_S$$\approx$$1.5$--$1.7$$\mu_B$\cite{han09,yaresko09}.    The moment measured by neutron scattering, in contrast, ranges 
from $0.25$$\mu_B$ (NdFeAsO)~\cite{kaneko08} to $1$$\mu_B$ (SrFe$_2$As$_2$)~\cite{chen08}.   
The AF ``stripe'' order found in Fe pnictides has also been observed in quasi-two dimensional insulating oxides with 
frustrated exchange interactions, where the ordered moment is strongly renormalised by quantum fluctuations\cite{skoulatos09}.  
By analogy, it has been suggested that magnetic excitations in Fe pnictides can also be understood in terms of a 
frustrated local-moment model,
\begin{eqnarray}
\mathcal{H}=J_{1x}\sum_{\langle ij \rangle_{1x}}\textsf{S}_i.\textsf{S}_j
+J_{1y}\sum_{\langle ij \rangle_{1y}}\textsf{S}_i.\textsf{S}_j
 +J_{1z}\sum_{\langle ij \rangle_{1z}}\textsf{S}_i.\textsf{S}_j
\nonumber \\
+J_2\sum_{\langle ij \rangle_{2}} \textsf{S}_i.\textsf{S}_j 
-K_{xy} \sum_i  \left((\textsf{S}_i^x)^2-(\textsf{S}_i^y)^2\right) + K_z \sum_i (\textsf{S}_i^z)^2
\label{eq:H}
\end{eqnarray}
where $\langle ij \rangle_{1\alpha}$ counts first-neighbor bonds in the $\alpha$-direction, 
$\langle ij \rangle_2$ second-neighbour bonds in the $x$-$y$ plane, and $K_{xy}$ is a single-ion anisotropy.   
It is interesting, therefore, to ask what constraints our two-fluid phenomenology places on this effective local-moment picture~?


A telling, and direct, comparison can be made in the context of the ordered moment.   
At a mean field level, the collinear ``stripe'' phase of the square-lattice \mbox{$J_1$--$J_2$} Heisenberg 
model,
\begin{eqnarray}
\mathcal{H}=J_{1}\sum_{\langle ij \rangle_{1}}\textsf{S}_i.\textsf{S}_j
+J_2\sum_{\langle ij \rangle_{2}} \textsf{S}_i.\textsf{S}_j ,
\label{eq:J1J2}
\end{eqnarray}
 becomes unstable for $J_2<|J_1|/2$
 \cite{shannon04}, or equivalently \mbox{$v_y<0$}.   Approaching this transition, 
quantum corrections to the ordered moment diverge, as illustrated in Fig.~\ref{fig:mzp}, and the sublattice magnetisation becomes zero before reaching the classical transition point.
For AF $J_1$, the dominant correction to $m_S$ comes from spin waves near the ordering vector.  
These are described by Eq.~(\ref{eq:S}) with $v_z = \Delta_\sigma = 0$, and we find, 
\begin{eqnarray}
\delta m_S = \frac{m_0}{2\chi_\perp}\frac{a^2}{(2 \pi)^2} 
\int_{|{\bf k}| < \Lambda} \frac{d {\bf k}}{\omega_{\bf k}}
= 
\frac{m_0 a^2\Lambda}{4 \pi^2 \chi_\perp v_x} {\sf K}_1(\kappa), 
\label{eq:elliptic}
\end{eqnarray}
where $\Lambda$ is a momentum cut-off reflecting the size of the spin-wave ``cone'', 
${\sf K}_1$ is a complete elliptic integral of the first kind, and \mbox{$\kappa = \sqrt{1-(v_y/v_x)^2}$}.  
At the limit of the $(\pi, 0)$ AF phase, $v_y \to 0$, and $\delta m_S$ diverges logarithmically\cite{chandra88}.   
The contribution of spin waves at higher energies must be determined separately,  
but for present purposes can be approximated by a constant offset $\approx 0.1 \mu_B$.


At first sight, fine-tuning a $J_1$--$J_2$ model into a region with $v_y \ll v_x$ offers the possibility of 
achieving any desired renormalisation of the ordered moment, $m_S$, cf.\cite{yao08,uhrig09,si08}.   
The same would hold of any itinerant electron model which could be mapped onto Eq.~(\ref{eq:S}).   
However, neutron scattering results for Fe pnictides suggest that $v_y \approx v_x$~\cite{diallo09}.  
Moreover, they clearly show a spin gap $\Delta_\sigma$, and out-of-plane dispersion $v_z$, both of which 
act to cut-off the divergence in $\delta m_S$.  


For a gapped, three-dimensional dispersion with \mbox{$v_y > \sqrt{v_xv_z(\Lambda/\pi)^3}$}, Eq.~(\ref{eq:S}) predicts
\begin{eqnarray}
\delta m_S 
&\approx&
 \frac{m_0 abc\Delta_\sigma}{8\pi^2\chi_\perp \bar{v}_s^3} \left( \Lambda \bar{v}_s \sqrt{1+\frac{\Lambda^2\bar{v}_s^2 }{\Delta_\sigma^2}} \right. \nonumber \\ 
&& \qquad \qquad \left. -\Delta_\sigma \ \mathrm{arcsinh}\left[ \frac{\Lambda \bar{v}_s}{\Delta_\sigma} \right]  \right), 
\label{eq:mzp}
\end{eqnarray}
where an energy cut-off $\epsilon=\Lambda \bar{v}_s$ has been imposed. 
For the purpose of comparison with experiment, the cut-off $\Lambda$ can be determined by the 
extent of the cone of linearly dispersing spin wave excitations seen in neutron scattering experiments.  
For parameters relevant to $\mathrm{BaFe_2As}_2$, where \mbox{$\Lambda \approx 0.2\pi/a$}~\cite{matan09}, 
Eq. (\ref{eq:mzp}) implies  $\delta m_S\approx 0.13\mu_B$, a value too small to explain the gulf between {\it ab intio} calculations
and experiment, although comparable with the smaller discrepancy with model based SDW theory~\cite{klauss08}.


In the highly frustrated region $v_y \to 0$, the approximation made in Eq.~\ref{eq:mzp} begins to break down, 
since the ellipsoidal integration region becomes longer and thinner, eventually escaping from the Brillouin Zone.  
Never the less Eq.~(\ref{eq:mzp}) does provide a finite bound, 
\begin{eqnarray}
\delta m_S < m_0\Lambda^3abc/(16\pi^2\chi_\perp\Delta_\sigma), 
\end{eqnarray}
on the maximum correction to the ordered moment from low-energy spin waves in a gapped, three-dimensional model.  
\footnote{Cylindrical choices of integration region provide better approximations for highly frustrated parameters.}

To illustrate how this works, in Fig.~\ref{fig:mzp} we compare the predictions
of the nonlinear sigma model, Eq.~(\ref{eq:S}), and the Heisenberg model, Eq.~(\ref{eq:H}), for 
the sublattice magnetisation, $m_S$, as a function of  $J_2$ --- and thereby $v_y$.  
Remaining parameters for Eq.~(\ref{eq:H}) are taken from experiments on $\mathrm{CaFe_2As_2}$~\cite{diallo09}.
Following LDA calculation\cite{han09}, we set the bare moment $m_0=1.51 \mu_B$.   
A constant offset $\delta m_S = - 0.3 \mu_B$ is added to Eq.~(\ref{eq:mzp}) to correct for high energy spin waves, 
and the value of $\Lambda$ is chosen so that the nonlinear sigma model predictions agree with the predictions of the
Hesienberg model at large $J_2/J_1$ (equivalently, large $v_y$). 
The agreement between these two approaches is excellent for a wide range of $J_2$.  
Even at the maximally frustrated point, the correction 
\mbox{$\delta m_S \approx 0.5 \mu_B \nonumber $}  
is smaller than the 
\mbox{$\delta m_S  \approx 0.7 \mu_B$}
needed to explain the discrepancy with experiment.  
We anticipate that this conclusion will hold for 
{\it any} spin model with realistic parameters\cite{schmidt10}, and conclude that the failure of LDA to accurately 
describe the size of the ordered moment lies in high-energy electronic correlation effects, 
not the zero point motion of low-energy spin waves. 

\section{Conclusion}

In conclusion, magnetic excitations in Fe pnictides are well-described by a simple two-fluid phenomenology 
in which gapped, three-dimensional spin waves co-exist with gapless, but incoherent particle-hole pairs.   
These two fluids can be treated as independent.  
This is evidenced by quantitative fits to NMR.
Our phenomenology is blind as to microscopic details of the real materials, but seems to fit
naturally with spin-density wave calculations which assign magnetism and metallicity 
to different, weakly coupled, sheets of the Fermi surface.

Furthermore it follows from explicit calculation that collective low energy spin fluctuations, of the type found in highly frustrated two dimensional quantum magnets, cannot be invoked to explain the  discrepancy in the LSDA and neutron scattering values for the sublattice magnetisation.


\acknowledgments
We are pleased to acknowledge helpful discussions with Andrey Chubukov, Amalia Coldea, 
Dima Efremov, Ilya Eremin, Cristoph Geibel, Stephen Hayden, Frederic Mila, 
Kenya Ohgushi, Toby Perring, Masashi Takigawa
and Alexander Yaresko. 
We endebted to Yasutomo Uemura for important conversations at an early stage of this work.
This work was supported under EPSRC grants EP/C539974/1 and EP/G031460/1

\end{document}